\documentstyle[art12, epsf]{article}

\begin{document}
\baselineskip 0.332in
\title{Light Gluino And Squark Production In Dijet Experiments 
At Tevatron}
\author{I. Terekhov and L. Clavelli\\
Department of Physics and Astronomy\\
The University of Alabama, Tuscaloosa Al 35487.}
\date {March 1996}
\begin{flushright}
UAHEP964\\March 1996\\hep-ph/9603390
\end{flushright}
\vskip 1cm
\begin{center}
{\Large Light Gluino And Squark Production\\[5mm] In Tevatron
Dijet Experiments}\\[15mm]
\def\thefootnote{\fnsymbol{footnote}}
{\large I. Terekhov \footnote{e-mail: ITEREKH3@UA1VM.UA.EDU}
and L. Clavelli\footnote{e-mail: LCLAVELL@UA1VM.UA.EDU}}\\[7mm]
{\it Department of Physics and Astronomy\\
The University of Alabama\\ Tuscaloosa Al 35487}
\end{center}
\vskip 1cm
\begin{abstract}
We consider single squark production in $p\overline{p}$ collisions at
$\sqrt{s}=1800$ Gev in the light gluino scenario, where production is
dominated by the parton subprocess 
$qg\rightarrow \tilde{q}\tilde{g}\rightarrow q\tilde{g}\tilde{g}$. Computed 
are the total production cross-sections as well as the contribution of such
production to the dijet mass distribution. 
We compare our cross-sections to the experimental data from the
CDF collaboration with integrated luminosity of $19 pb^{-1}$.
\end{abstract}
\thispagestyle{empty}
\setcounter{page}{0}
\newpage

\section{Introduction}

\indent

Among the new particles predicted by Supersymmetry \cite{SUSY}
are strongly interacting scalar quarks and gluinos. In the experimental
searches for the new particles multiple constraints on allowed mass regions
have been imposed. While many lower limits on the gluino mass have been
discussed in the literature \cite{litediscuss}, 
there still remains a possibility of an
ultra-light (below 1 Gev) gluino \cite{ua1}. The high energy colliders
have been rather insensitive to this narrow region of the gluino mass. On
the other hand, a (nearly) massless gluino may even be favored by the present 
experimental data, including $\alpha_s(M_Z)$ measurements \cite{alphas}
and those on
$Z\rightarrow b\overline{b}$ decay \cite{Zbb}.

The light gluino case also provides somewhat better grounds for the scalar 
quark search.
In this scenario, a single massive squark with a (nearly) massless gluino
can be produced via $qg\rightarrow \tilde{q}\tilde{g}$. Such a reaction
would considerably dominate similar reactions of two heavy particle
production (e.g. 
$q\overline{q},\, gg\rightarrow \tilde{q}\tilde{\overline{q}}$)
such as would be required in the heavy gluino case.
Furthermore, the experimental signature of such heavy particles is fairly
complicated, possibly involving multijet events. On the contrary, in the light 
gluino case the scalar quark would decay mainly into a quark and a
gluino. Although the hadronization of a  gluino is not well known, it
is reasonable to assume that a light gluino primarily results in a single jet,
thus allowing squark production to reveal itself simply as a peak in
the dijet mass distribution. In this article we suggest how the 
experiments on dijet mass distribution, which are
in particular being performed by the CDF at the Tevatron, can search for
scalar quarks in the $200-800$ Gev mass range. In this work the reaction
$ug \rightarrow \tilde{u}\tilde{g} \rightarrow u\tilde{g}\tilde{g}$
is studied in light of the CDF experimental conditions. 

\section {Production Cross Sections}
\indent

\begin{figure}
\hskip 3.5cm
\epsfxsize=3in \epsfysize=3.5in \epsfbox{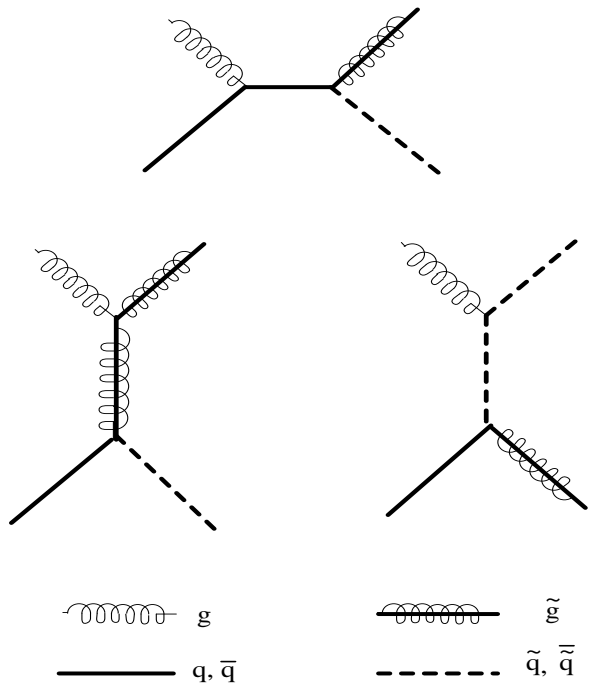}
\caption{Feynman graphs for single squark production.}
\label{two2two}
\end{figure}

The Feynman graphs for the single squark production 
$qg \rightarrow \tilde{q}\tilde{g}$ are shown in Fig. 1.
Expressions for the squared amplitude on the parton level can be found in 
\cite{m2src1,m2src2}. On discarding terms vanishing in the limit 
$m_{\tilde{g}}=0$ the cross-section reads \[
\hat{\sigma}_{tot} = 
\frac{\pi\alpha_s^2}{\hat{s}}
\{ (1+2\mu^4-2\mu^2)\log{\frac{(1-\mu^2)^2}{m^2}}
+ \frac{1}{18}(1-\mu^2)(65\mu^2-14) 
+ \frac{1}{18}\mu^2(16 - 19\mu^2)\log{\mu^2}
\},
\]
where $\mu^2=M_{\tilde{q}}^2/\hat{s}, m^2=m_{\tilde{g}}^2/\hat{s}$.
This result disagrees with \cite{m2src2}.
We use CTEQ 3L (leading order QCD fit) parton densities \cite{tung}. 
$\alpha_s(Q), q(x, Q), g(x, Q)$ are evaluated at $Q=\sqrt{\hat{s}}$.
We also sum over left/right squark production neglecting the possible mass 
splitting between these states.
Since there is no interference between diagrams containing $\tilde{q}_L$ and 
$\tilde{q}_R$, in the case of $M_R-M_L$ mass splitting large relative to the
$\tilde{q}$ width one simply divides the production cross-sections by 2 to
get separate production cross-sections.
We have taken the gluino mass to be 100 Mev.
The production cross-sections are plotted as a function of squark mass
in Fig. 2.

\begin{figure}
\hskip 2.5cm
\epsfxsize=4in \epsfysize=3in \epsfbox{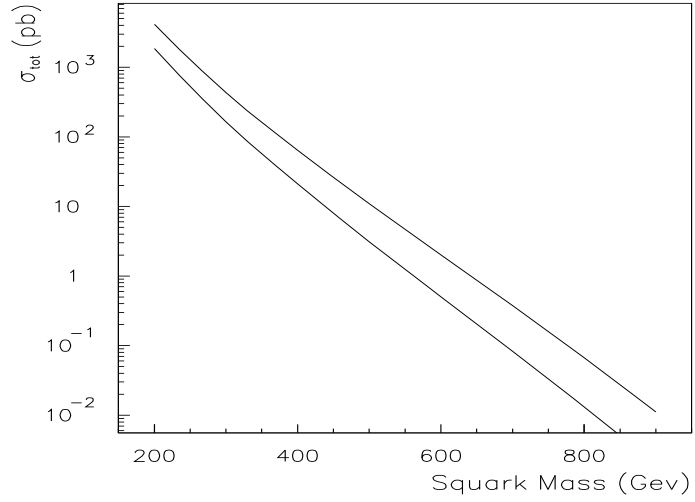}
\caption{Single squark production cross-sections for 
$\tilde{u}$ (upper curve) and $\tilde{d}$; $m_{\tilde{g}}=100$ Mev.}
\label{two}
\end{figure}

\section {Dijet Cross Sections}
\indent

\begin{figure}
\hskip 3cm
\epsfxsize=4in \epsfysize=2.6in \epsfbox{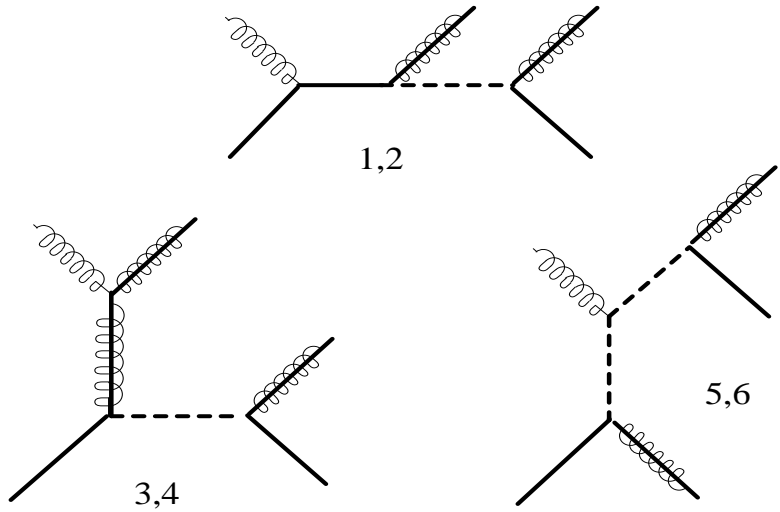}
\caption{Resonant terms.}
\label{res}
\end{figure}

For the purposes of making experimental
predictions, however, as well as for the shape of the peak and dijet angular
distribution, one has to consider the pure final state
$q\tilde{g}\tilde{g}$. The corresponding Feynman graphs are shown
in Fig. 3. In the light gluino case considered here
squark production is dominated by the gluino exchange graphs 3,4 in
Fig. 3, which lead to a predominantly forward gluino and hence
to a two jet topology.
In addition to the square
of the resonant terms we consider their interference with the
non-resonant amplitudes from Fig. 4.
We do not include the square of non-resonant terms since it is a 
part of the background as long as we  study only
the resonances in the dijet mass distributions rather than complete
SUSY contribution into dijet cross-sections. 
The interference terms, however, prove to be rather insignificant.

\begin{figure}
\hskip 3cm
\epsfxsize=4in \epsfysize=3in \epsfbox{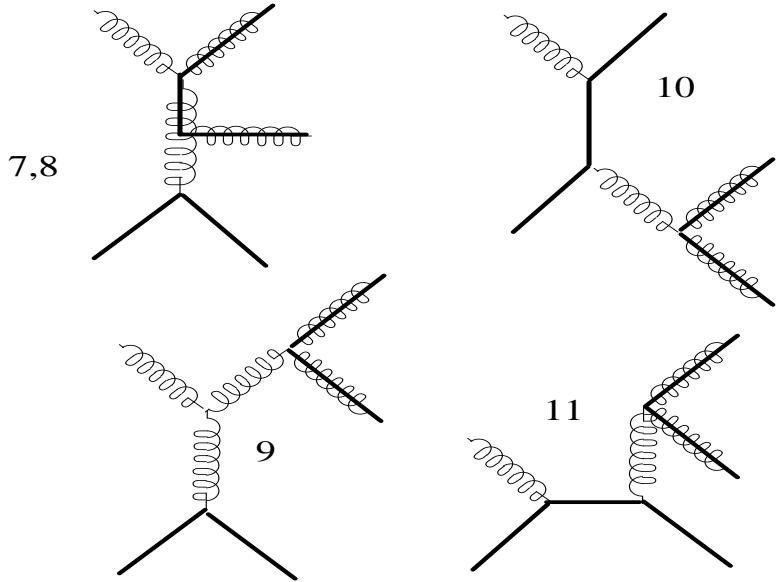}
\caption{Nonresonant terms.}
\label{nores}
\end{figure}

Dijets as studied at the CDF are defined as pairs of jets $a,b$ with highest
$p_t$ in the events which pass certain selection criteria as well as the
pseudorapidity cuts
$\left | \eta_{a,b} \right | < 2$, or
$\left | \cos\theta_{a,b} \right | < 0.964$, and
$\left | \tanh\frac{\eta_a-\eta_b}{2} \right | < 2/3$, or
$$1/5 < \sqrt{\frac{(1-\cos\theta_a)(1+\cos\theta_b)}
                        {(1-\cos\theta_b)(1+\cos\theta_a)}} < 5.$$

Since the CMS angle of a dijet with zero net transverse energy is
$\cos\theta^*= \tanh\frac{\eta_a-\eta_b}{2}$, the last cut
is used in the experiments as the cut on $\cos\theta^*$
to provide a uniform detector acceptance as well as to reduce standard QCD
background \cite{main}.
We compute the part of dijet mass distribution due to the scalar quark
production as

\begin{eqnarray*}
\lefteqn{
\Delta \frac{d\sigma(p\overline{p}\rightarrow 2jets + X)}{dM}=
\frac{d\sigma_{with\, \tilde{q}}-d\sigma_{without\, \tilde{q}}}{dM} }  \\
& & = 2\int dx_1\, dx_2\, dLIPS\, \frac{1}{2\hat{s}}\left | 
\overline{{\cal M}^2 } \right | q(x_1, Q) g(x_2, Q)F.
\end{eqnarray*}
where the factor of 2 accounts for  antisquark production, 
$ {\cal M}^2 = \sum_{i,j=1}^{11}{{\cal M}_i{\cal M}_j^*} 
- \sum_{i,j=7}^{11}{{\cal M}_i{\cal M}_j^*}
$
and\[
F=\theta(p_{5\bot}-p_{3\bot})\theta(p_{4\bot}-p_{3\bot})
  \delta(\sqrt{(p_4^\mu+p_5^\mu)^2} - M) + cyclic \;perm.,
\]

\begin{figure}
\hskip 3cm
\epsfxsize=4in \epsfysize=3in \epsfbox{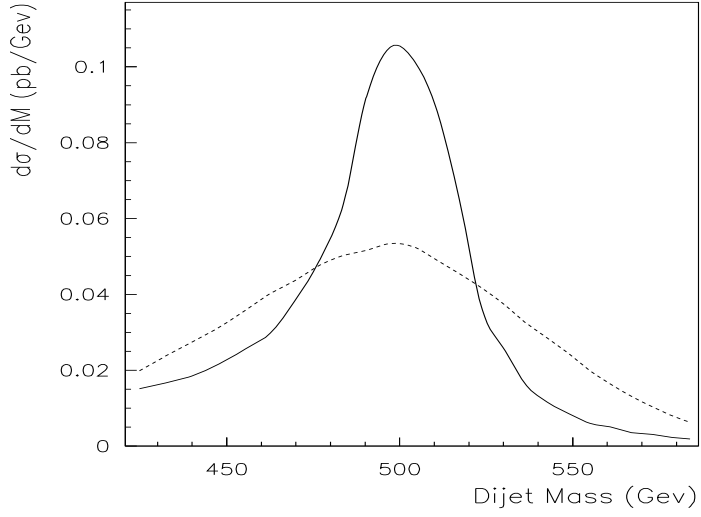}
\caption{Squark resonance in dijet mass distribution.}
\label{shape}
\end{figure}

The total squark width is approximated by its ``hadronic'' width
$ \Gamma(\tilde{q}\rightarrow q\tilde{g}) =2/3\alpha_s m_{\tilde{q}}
\approx \Gamma_{tot}$; we neglect other decay channels.
The ratio of this width to the squark mass
is smaller than the mass resolution (10\%) at the CDF \cite{main}, which
is crucial for the reaction being detectable.

We perform a six-dimensional Monte-Carlo integration taking into account
all applicable experimental cuts. A typical resonance, both pure and smeared
with 10\% detector resolution is depicted in
Fig. 5, where we have assumed a squark mass of 500 Gev. 
The total resulting cross-section as a 
function of squark mass is plotted in fig. 6. One can see that,
as can be expected from the cuts, the dijet cross-section is roughly $0.6-0.7$
of the production cross-section shown in Fig. 2.
Scalar quarks in this mechanism are mainly produced near 
$\left | \cos \theta \right | = 1$ and decay isotropically in their rest 
frame. Since $\eta_a-\eta_b$ is invariant under Lorentz boosts along the
beamline, the distribution in $\cos \theta^*$ is nearly flat and not
shown.

\section{Conclusions}

\indent

\begin{figure}
\hskip 3cm
\epsfxsize=4in \epsfysize=3in \epsfbox{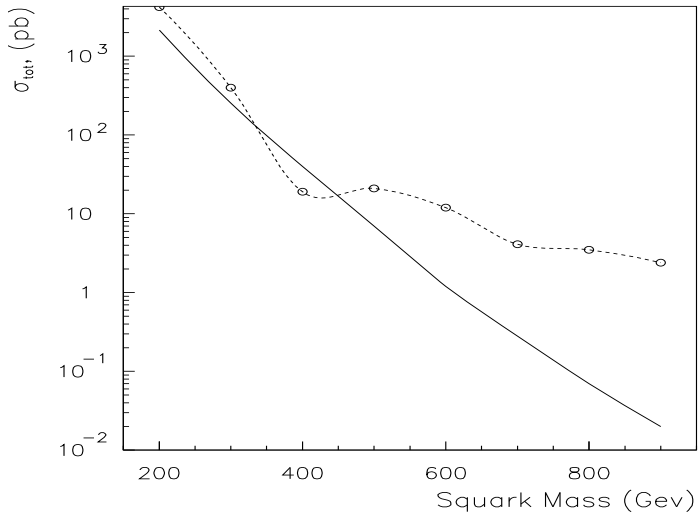}
\caption{Total dijet cross sections due to $\tilde{u}$
production. Also shown (dashed curve) is CDF 95\% CL upper limit on the
cross section times the branching ratio for new particles decaying into
dijets.}
\label{three}
\end{figure}

We compare the predicted cross-sections with the CDF data on 95\% CL
upper limits on new narrow resonances \cite{main}. 
Our analysis indicates that 
a left-right degenerate $\tilde{u}$ in the mass region of $330-440$ Gev is 
apparently  inconsistent with the light gluino scenario. 
One can see that the dijet cross-sections
associated with single squark production are large enough to bring about 
discernible signals in the broad mass region of $200-700$ Gev. 
However, the CDF sensitivity to new particles , that is, the position of 
the exclusion curve in Fig. \ref{three}
is also determined by the systematic errors as well as
by large prescaling factors for triggers \cite{main,pc}. 


In the light gluino scenario, jet experiments should prove to be a viable tool
in the search for massive scalar quarks, as we have demonstrated in connection
with dijet mass distributions. Current CDF data on the dijet mass spectrum 
appears to exclude $\tilde{u}$ squarks of mass between 330 and 440 Gev if the 
gluino is light. With slight improvement of the experimental 
sensitivity, scalar 
quarks below 800 Gev will be observed or proven inconsistent with a light 
gluino. In the standard SUSY picture, where squarks and gluinos are heavy,
squark production cross-sections \cite{heavy} 
are orders of magnitude below those
computed here and dijet decays are not expected. A squark of
mass $M$ will also in the light gluino scenario produce a Jacobian peak
in the jet transverse energy distribution at approximately $M/2$. We will
address this issue in a forthcoming paper.

\vskip 1cm
{\Large \bf \noindent Acknowledgements}
\vskip 1cm

We would like to thank Dr. R. Harris from the CDF collaboration at Fermilab
for indispensable information about the CDF experiment. Our numerical 
integrations were performed using the Gottschalk Monte-Carlo program.
This work was supported in part by the Department of Energy under 
grant DE-FG05-84ER40141.

\end{document}